> …when I die and go to Heaven, there are two matters on which I hope for enlightenment. One is quantum electrodynamics, and the other is the turbulent motion of fluids. And about the former I am really optimistic.
>
> Sir Horace Lamb, 1932

# Mechanical models of physical fields and particles

**Valery P. Dmitriyev**

*Lomonosov University*
*P.O.Box 160, Moscow 117574, Russia*
*e-mail: dmitr@cc.nifhi.ac.ru*

**Abstract**

Earlier obtained results on mechanical analogies of physical fields and particles are reviewed. The approach rests on the concept of the substratum – a mechanical medium, which occupies all the space and serves as a seat to support the light and to transmit interactions. A turbulent ideal fluid was chosen for the substratum. The turbulence is supposed to be homogeneous and isotropic in its ground state. Perturbations of the turbulence model physical fields. Particles originate from the voids in the fluid. Symmetrical pairs of particle-antiparticle find analogies in mechanical pairs of cyclone-anticyclone. A quantum particle is modeled by the dispersion of a point discontinuity (defect) in the stochastic medium. Gravitation relates to emitting by defects the continual flow of the transient point dilatation. The shock wave mechanism of the re-collection a discontinuity in the incompressible medium governs such phenomena as the "wave function collapse" and instantaneous quantum correlations. Microscopically, the electromagnetic wave and gravitation are modeled by the torsional and axisymmetric waves, respectively, in the vortex sponge.

PACS numbers:
12.60.-i – Models of particles and fields beyond the standard model.

## 1. Physics reduced to mechanics

Much indicates that the sectors of physics among those, which are traditionally regarded as nonmechanical, can be nevertheless expressed in terms of continuum mechanics. The object pursued is to include in classical mechanics as much of physics as possible reaching in this the extremes. One may expect then that the rest of physics are apt to be covered by a consistent "analytical extension" of the mechanics. In this event, there should be used additional postulates capable to generalize it inartificially.

As a matter of fact, such the additions are already present in first approaches to expand the traditional areas where classical mechanics is applicable. Thus, when constructing the mechanical models of electromagnetic fields, we deal with firstly stochastized and then averaged mechanics of turbulent media. Maxwell's equations appear to be isomorphic to some subset of the model, complying in particular with the condition of that the higher fluctuation moments getting vanished [1, 2]. In the bounds of the mesoscopic model the latter supposition is actually a postulate or a hypothesis whose validity can be verified only via the consideration of a microscopic mechanism.

Schroedinger equation corresponds to the law of dispersing and (hyper)diffusion a discontinuity of a turbulent medium. When deriving it in the point model of a defect, it is impossible to do without imposing on second derivatives the restriction, which is a continuous generalization of the classical mechanics postulate regarding the equality of the acceleration to a function of space variables [3]. As well, Schroedinger equation can be extracted from a microscopic model [4].

The discreteness of the mass and charge is modeled with the aid of the postulate, whose origination from mechanics is not evident. Specifically, the reciprocity is admitted for interaction of the stress sources in the medium (the Betti's "theorem"). The latter is not a corollary of Newton's mechanics. Otherwise, the uniqueness of the vortex structure underlying should be substantiated [1].

## 2. Substratum for physics

A consistent mechanical model of physical phenomena deals with the motion of a continuum, which represents a substratum for physics [5]. In this event, the light corresponds to the perturbation wave in the substratum, physical fields – to its bulk disturbances or deformations. A substratum defines the absolute frame of reference. Anyway, for small disturbances it is not distinguished among other inertial frames.

## 3. Relativity

The relativity follows immediately from the substratum approach. As a rough model of the substratum one may take a solid elastic body liable to sheer deformations [5]. The transverse modes of this medium model the electromagnetic phenomena. The lack of the longitudinal waves implies the incompressibility of the substratum. It is the latter that conditions the instantaneity of the "wave function collapse" and that of the gravity. The motion of the solid body substratum is governed by d'Alembert equation, which is known to be Lorentz invariant. Thus, the substratum model supplies a microscopic mechanism of relativity. Relativistic kinematics finds its explanation as a dynamics of the mechanical substratum. In other words, relativity turns out to be a phenomenology with respect to substratum models. Therefore, it will be a more correct terminology to speak about the relativistic model of the particle's high-speed motion, on the one side, and its mechanical-substratum theory, on the other.

## 4. The hydrodynamic substratum

An adequate model of the substratum is provided by a turbulent ideal fluid with voids. Here, ideal means inviscid, incompressible, noncorpuscular. Perturbations of the turbulence model electromagnetic fields [6]. The voids relate to particles of matter [1].

Formally, the general perspectives and expectations over the fluid model are based on the following considerations. Due to relatively loose character of the pressure term and the additional degrees of freedom introduced by the inclusions of the empty space, the hydrodynamic equations appear to be very rich in their solutions. With this, the appropriate choice in boundary conditions enables us to obtain the structures required.

Someone may argue the very existence of turbulence in frictionless fluids. However, there are no unambiguous evidences of a nonclassical origin of the turbulence [7] and cavitation [8] observed in superfluids. Anyway, in the present context the notion of turbulence is employed [1, 2] as a large-scale description of a fine vortex structure [4].

## 5. Turbulence averaged

The fluid motion is specified by the flow velocity $\mathbf{u}(\mathbf{x},t)$, pressure $p(\mathbf{x},t)$ and volume density $V(\mathbf{x},t)$. In a developed turbulence these quantities can be regarded as random functions of the space $\mathbf{x}$ and time $t$ variables. The respective averages $\langle \mathbf{u} \rangle$, $\langle p \rangle$ and $\langle V \rangle$ are already the regular functions of $\mathbf{x}$ and $t$. Centering the random quantities we consider the turbulence fluctuations

$$\mathbf{u}' = \mathbf{u} - \langle \mathbf{u} \rangle, \qquad p' = p - \langle p \rangle .$$

The volume density is supposed to not fluctuate i.e.

$$V = \langle V \rangle .$$

Following this approach, the turbulence can be described on the average by the series of the point moments $\langle u_i \rangle$, $\langle p \rangle$, $\langle V \rangle$, $\langle u'_i u'_k \rangle$ etc. The respective set of differential equations is found through manipulating and averaging the Euler equation, the latter describing the motion of an ideal fluid. This just constitutes the Reynolds technique commonly used in hydrodynamics in order to handle with the developed turbulence. Further, we break the infinite chain of Reynolds equations supposing that the third moments are vanishing, as is the case in the absence of discontinuities. In general, the remainder term can be expressed via the volume density of the point discontinuities in the medium. The closure to the set of the equations thus obtained provides us with an approximate description of the averaged turbulence.

## 6. The high-energy low-pressure substratum

The second moments of the velocity fluctuations have the meaning of the turbulence energy volume density

$$1/2 V \langle u'_i u'_i \rangle . \qquad (6.1)$$

Here and further on the summation over recurrent is implied throughout. It is convenient to designate

$$p = \mathbf{p}/V_0 .$$

In the ground state the turbulence is supposed to be homogeneous and isotropic i.e..

$$\langle V \rangle^0 = V_0 = \text{const}, \quad \langle \mathbf{u} \rangle^0 = 0, \quad \langle p \rangle^0 = \text{const}, \quad \langle u'_i u'_k \rangle^0 = c^2 \mathbf{d}_{ik}, \qquad (6.2)$$

where $c = \text{const}$. The special case of the turbulence is further considered – with the relatively low pressure and high energy density:

$$\langle p \rangle^0 \ll c^2 , \qquad (6.3)$$

and also

$$\langle p \rangle \geq 0 .$$

This model configuration will be chosen [1] as a substratum for physics.

## 7. Maxwell's electromagnetic equations

We are interested in deviations from the background (6.2)

$$\mathbf{d}\langle \mathbf{u} \rangle = \langle \mathbf{u} \rangle - \langle \mathbf{u} \rangle^0,$$
$$\mathbf{d}\langle p \rangle = \langle p \rangle - \langle p \rangle^0,$$
$$\mathbf{d}\langle u'_i u'_k \rangle = \langle u'_i u'_k \rangle - \langle u'_i u'_k \rangle^0 ,$$
$$\mathbf{d}\langle V \rangle = \langle V \rangle - \langle V \rangle^0 .$$

Supposing that

$$\left| \mathbf{d}\langle p \rangle \right| \leq \langle p \rangle^0 ,$$

we find from (6.3) and the Reynolds equations implied the following evaluation:

$$\left| \mathbf{d}\langle p \rangle \right| \ll c^2 , \quad \left| \mathbf{d}\langle u'_i u'_k \rangle \right| \ll c^2 , \quad \left| \mathbf{d}\langle u \rangle \right| \ll c , \qquad (7.1)$$

and also

$$\left| \mathbf{d}\langle V \rangle \right| \ll V_0 .$$

In this way the Reynolds equations can be linearized. As being closed straightforwardly, the set of the linearized Reynolds equations appears to be isomorphic to Maxwell's electromagnetic equations. In this event, the following correspondence takes place [6]:

$$A_i = \kappa c \delta \langle u_i \rangle, \tag{7.2}$$

$$\mathbf{j} = \kappa \delta \langle p \rangle, \tag{7.3}$$

$$E_i = \kappa \P_k \delta \langle u'_i u'_k \rangle,$$

$$j_i = \frac{\kappa}{4\pi} \partial_k h_{ik},$$

where $\kappa$ is an arbitrary constant, $h_{ik}$ – is a remainder term involving the third moments of the fluctuations, and $\pi$ – the Euler number.

Notice also, that (in the incompressible substratum} the spreading of the disturbance wave, modeling the light, is associated solely with the variation of the nondiagonal (cross) moments. I.e. in this process the turbulence energy density does not vary [1].

## 8. Cavitons

The neutron is modeled by a cavity in the turbulent fluid. Such a configuration is unstable. In the case of a molecular liquid the cavity fills with the vapor and thus the thermodynamic equilibrium is established. If the fluid does not consist of corpuscles i.e. it is a true continuum, then the mechanical balance is established by means of a deviation of the turbulence from the ground configuration up to dropping the pressure at the cavity surface to zero. That stable cavity together with the turbulence perturbation field produced by it models the proton and the corresponding electrostatic field (fig.1). Indeed, it can be shown [1] that

$$\delta \langle u'_1 u'_1 \rangle = \frac{\langle p \rangle^0 R}{|\mathbf{x}-\mathbf{x}'|}, \tag{8.1}$$

where $R$ is the cavity radius with the center in $\mathbf{x}'$. For electrostatic field the full energy of turbulence is infinite. Hence, the positive deviations of the energy density from the background must be compensated with the respective negative deviations in another place. This leads to an idealized model of the electron as the inclusion of an isle of the quiescent fluid (fig.3):

$$\delta \langle u'_1 u'_1 \rangle = -\frac{c^2 r_e}{|\mathbf{x}-\mathbf{x}'|}, \tag{8.2}$$

$$c^2 r_e = \langle p \rangle^0 R, \tag{8.3}$$

where $r_e$ is the radius of the inclusion.

A mechanism for generating the pair of opposite electric charges involves as a starting point the system of two nonequilibrium cavities in a turbulent fluid. For instance, it may be a large (neutron) and a little (neutrino) cavities. When approaching the equilibrium, the small cavity shrinks in, forming an isle of the quiescent fluid (the electron, fig.3). The large cavity expands accordingly, forming a stable inclusion of the empty bubble (the proton, fig.1). Such does the cavitation proceed in a discontinuous medium. The stable formations resulted combine in itself both the dilatational inclusions and the sources of stationary turbulence perturbation field. Those perturbation centers are referred to as *cavitons* [1]. In the point representation, a dilatational inclusion is modeled by the center of dilatation

$$\nabla \cdot \mathbf{s} = \Delta V \, \delta(\mathbf{x} - \mathbf{x}'), \tag{8.4}$$

where $\mathbf{s}$ is the medium displacement. A bubble expanded corresponds to $\Delta V > 0$, the one shrinked in – to $\Delta V < 0$.

## 9. Electrostatic interaction

In the medium stressed with the pressure the dilatation performs the work

$$\begin{aligned} E &= V_0 \int \mathbf{s} \cdot \nabla \, dp \, d^3x \\ &= -V_0 \int dp \, \nabla \cdot \mathbf{s} \, d^3x \, . \end{aligned} \tag{9.1}$$

Thus, the gradient pressure field exercises a force on the dilatational inclusion. The force affecting between the two cavitons models the Coulomb interaction of the two charges. From (9.1), (8.4) and with account of (7.3), we have [1] the particle definition of the electric charge:

$$q = -V_0 \Delta V / k \, . \tag{9.2}$$

## 10. The Lorentz gauge

According to (7.2), the Coulomb gauge corresponds to the condition of the medium incompressibility

$$\nabla \cdot d\langle \mathbf{u} \rangle = 0 \, .$$

The Lorentz gauge comes from the continuity equation, which is linearized due to the minuteness of the medium compressibility:

$$\partial_t d\langle V \rangle + V_0 \nabla \cdot d\langle \mathbf{u} \rangle = 0 \, .$$

The relation required is obtained [2] putting in the latter the Hooke's law

$$V_0 d\langle p \rangle = c^2 d\langle V \rangle$$

and taking into account the definitions (7.2), (7.3).

## 11. Particles and antiparticles

Formally, the condition of medium compressibility enables us to get for cavitons symmetrical solutions, which correspond to particle-antiparticle pairs [2]. See, for instance, the models of the proton and antiproton in the compressible medium (fig.1, 2). In a meteorological sense, those symmetries correspond to cyclone-anticyclone pairs.

However, basically, we must indicate the microscopic mechanism for the compressible medium capable to support that sort of configurations. Such a mechanism is provided by the vortex sponge – a disorderly heap of the hollow vortex tubes, which pierce the fluid in all directions. On a large scale, the vortex sponge looks as a homogeneous dispersion of the empty space in a microscopically continuous medium. Singular leaps of the density relate to solitons of the vortex sponge [4]. For instance, in the case of the neutron/proton that can be a loop on the vortex filament. When moving, a loop takes the features of the vortex ring.

## 12. Two kinds of the energy

One must distinguish the energy of "vacuum" and the energy of physical field [1, 4]. Vacuum energy corresponds to general energy of the turbulent motion of the substratum (6.1). The energy of electromagnetic field is represented by the integral of Maxwell's equations. In the bounds of the linear model, the energy of field does not intermingle with the energy of vacuum. Microscopically, the energy of electromagnetic field corresponds to the energy of bending the vortex filament.

## 13. Discreteness of the cavitons

Postulating identity of the field (8.1)-(8.3) and particle (9.2) definitions of the electric charge, we get:

$$\boldsymbol{k}c^2 r_e = V_0 |\Delta V|/\boldsymbol{k} . \qquad (13.1)$$

On the other side, it can be assumed for the dilatation

$$|\Delta V| = \frac{4\boldsymbol{p}}{3} r_e^3 .$$

Putting this to (13.1), we get an equation to find a unique radius of the caviton.

The above convention is equivalent to that the reciprocity of the elastic interaction (9.1) is valid in the medium under consideration [1]. The latter requirement represents an additional postulate, which extends the bounds of Newton mechanics.

## 14. Dispersing a caviton

Notice that the idealized model of the electron (fig.3) violates the requirement of smallness (7.1) for turbulence perturbations. For this reason, the respective caviton occurs in the turbulent medium as a delocalized formation (fig.5). In such a configuration, the radius of the isle of the comparatively quiet fluid $R_e$ appears to be equal or even larger than the radius of the stable cavity $R$:

$$R_e \approx 1.5R,$$

that is just the so called classical radius of the electron $R_e$ [1].

A delocalized caviton can also be viewed [2] as composed of fragments each of those satisfies the condition (7.1). Take notice that the radius of the fragment (fig.4) equals to the radius of the localized caviton (fig.3). Splitting and distributing of a caviton in the turbulent medium leads to its dispersion being formed. The latter is a kind of the gas – a gas without the hydrostatic pressure. The "particles" of this gas possess different masses although have one and the same size of the core. Two "particles" of a kind are genuine indistinguishable from each other. An element of the caviton's dispersion loses its authenticity just in the next moment of evolution i.e. it is never identical to his earlier version. Therefore, a caviton has no trajectory of motion in principle. All this enables us to suggest that the dispersion of a point discontinuity in the turbulent medium (the dispersion of a singularity) represents a mechanical implementation of the Madelung medium [3, 5].

## 15. Gravitation

The mechanism of gravitation bears on the contact interaction of the dilatation centers, which each defect of the medium is supposed to emit [5, 9]. Unlike the cavitons, those dilatation centers are nonstationary and capable to exist only in dynamics. Microscopically, the flow of the transient point dilatation corresponds to the axisymmetric wave spreading along a vortex tube. The speed of this wave $c_g$ exceeds enormously the speed of the torsional wave $c_r$ propagating as well along the vortex filament. The torsional wave provides a model of the electromagnetic wave or the photon. In its turn, the axisymmetric wave corresponds to the wave of gravitation.

On the whole, we have the interrelation between the speeds of the longitudinal and transverse waves, respectively, which is typical for a jelly:

$$c_g >> c_r.$$

Ideally, when the medium is incompressible, the longitudinal wave spreads instantaneously

$$c_g \to \infty.$$

This factor determines the shock wave or avalanche mechanism of the re-collection a discontinuity in the incompressible medium, which is responsible for such phenomena as the "wave function collapse" and instantaneous quantum correlations [5, 3].

## 16. Conclusion

It can be stated in the end that a turbulent ideal fluid with voids provides a suitable medium for constructing a mechanical analogy of physical fields and particles. On the whole, the following isomorphism between the notions of physics and those of continuum mechanics takes place (see the table). The completeness of the analogy thus obtained enables us to speak of a mechanical (or semimechanical) model of physical fields and particles.

Concerning the general perspectives for the approach, the present contribution should be regarded as a first step only and even a tentative one. The next stage must be to render the continuum mechanics underlying in terms of the gauge fields. And to see to what extent it coincides with the existent theory and what may be the differences.

| concept of physics | property of a turbulent fluid |
|---|---|
| electromagnetic wave | turbulence perturbation wave |
| magnetic vector-potential | average velocity of the fluid flow |
| electrostatic potential | average fluid pressure |
| electromotive force | difference in the density of the average turbulence energy |
| formation of the pair $(p^+, e^-)$ of opposite electrical charges | turbulent cavitation of a discontinuous fluid |
| electrical current | flow of the cavitational dilatation |
| mass of a particle | equivalent mass of a cavity in the fluid |
| particle | cyclone |
| antiparticle | anticyclone |
| Madelung gas | dispersion of a point discontinuity |
| collapse of a particle | avalanche of disruption |
| gravitation | flow of the transient point dilatation |

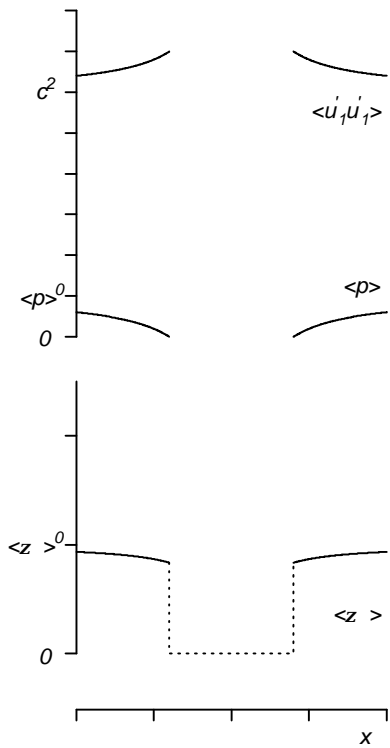 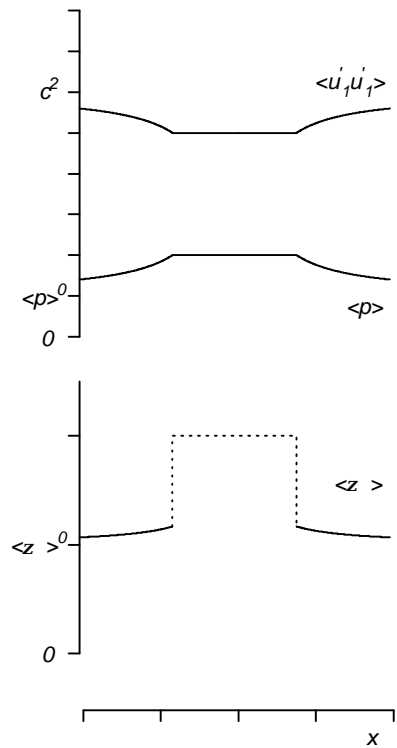

Fig. 1                                       Fig. 2

The hollow-bubble model of the proton.         The antiproton.

.

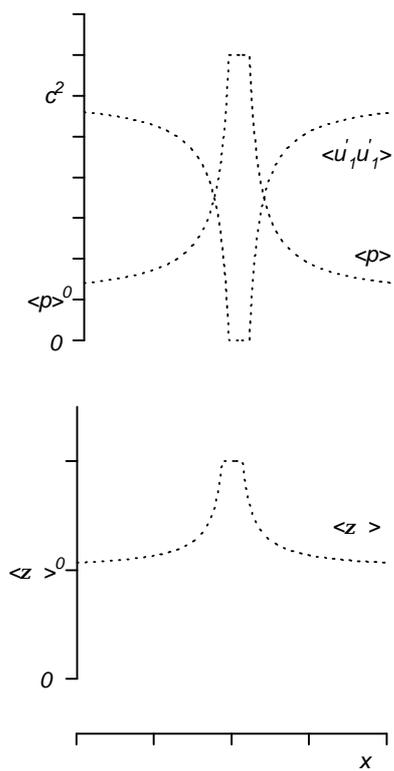
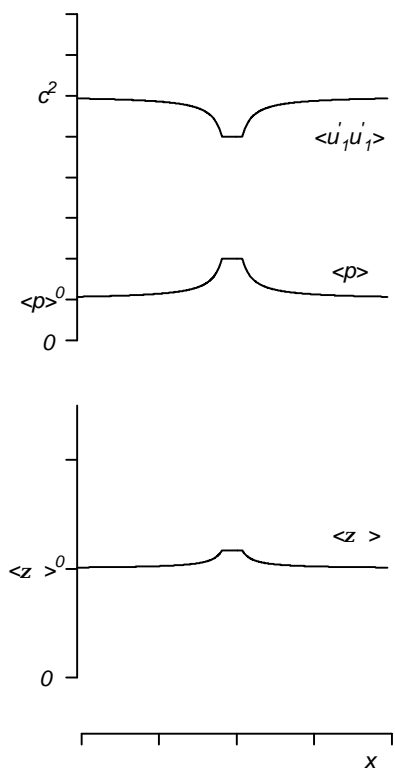

Fig.3

Fig.4

The localized electron.

The $1/N$-th splinter of the electron, $N = c^2 / \langle p \rangle^0$. Here $N=6$.

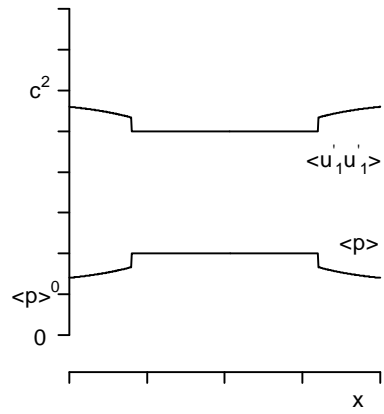

Fig. 5

The negative caviton (Fig.3, top) flattened to the level of the wave amplitude.